\documentclass[mathleft
]{an}
\usepackage{graphicx}
\usepackage{times}
\begin{document}

\Pagespan{789}{}
\Yearpublication{2006}%
\Yearsubmission{2005}%
\Month{11}%
\Volume{999}%
\Issue{88}%

\title{The population of ULXs in the spiral galaxy NGC~2276}

\author{Anna Wolter\inst{1}\fnmsep\thanks{Corresponding author:
  \email{anna.wolter@brera.inaf.it}\newline},
Fabio Pizzolato\inst{1},
Stefano Rota\inst{2}, 
Michela Mapelli\inst{2}, \and  Emanuele Ripamonti\inst{2}
}
\titlerunning{ULXs in NGC~2276}
\authorrunning{A. Wolter et al.}
\institute{
INAF-Osservatorio Astronomico di Brera
Via Brera, 28 
20121 MILANO
ITALY
\and 
Universit\`a degli Studi Milano-Bicocca,
MILANO
ITALY
}

\received{}
\accepted{}
\publonline{later}


\abstract{
We present results for X-ray  point sources in the Sc galaxy NGC~2276, obtained
by analyzing Chandra data. The galaxy is known to be very active in many 
wavelengths, possibly due to gravitational interaction with the central 
elliptical of the group, NGC~2300. 
However, previous XMM-Newton observations resulted in the detection of
only one bright ULX and extended hot gas emission. We present here the
X-ray population in NGC~2276 which comprises 17 sources. We found that
6 of them are new ULX sources in this spiral galaxy resolved for the
first time by Chandra. We constructed the Luminosity Function 
that can be interpreted as mainly due of High
Mass X-ray binaries, and estimate the Star Formation rate (SFR) to be 
SFR $\sim 5-10 M_{\odot}$/yr.
}

\maketitle

\section{Introduction}
\label{Intro}

NGC~2276 is a peculiar Sc galaxy belonging to the group of NGC~2300, the first
small group of galaxies in which hot diffuse intragroup gas has been observed
in X-rays with ROSAT (Mulchaey et al. 1993, Davis et al. 1996). The mass 
estimate 
of the hot gas is M$_{gas} = 1.25 \pm 0.13 \times 10^{12} {\rm M_{\odot}}$.
The X-ray emission, of about 0.3 Mpc extent, is centered close but not
exactly on the central elliptical NGC~2300. 
The emission is well described by a King model
with a core radius r$_{o} = 4.28 \pm 1$ arcmin and $\beta  = 0.41\pm 0.3$
and has a luminosity of L$_X^{bol} = 1.12 \times 10^{42}$ erg/s
(Davis et al. 1996).

The peculiar ``bow shock'' shape of NGC~2276 underlines the interaction with
either the hot gas surrounding the group center or directly with the
central elliptical. Condon (1983) suggests the presence of ram-pressure 
interaction between NGC~2276 and the IGM, while Gruendl et al. (1993) and 
Elmegreen et al. (1991) explore tidal interaction with the central
elliptical galaxy NGC~2300.
Davis et al. (1997) note that data in all observed wavelengths show
enhancement of emission in the SW quadrant. They derive that the morphology
of the galaxy is primarily affected by the gravitational encounter which
has disturbed both the stellar and gaseous component and enhanced
star formation rate along the truncated side of the galaxy.
They conclude that ram pressure may enhance the star formation, but is not the 
dominant force shaping the galaxy.

This Sc galaxy is indeed very active. Many supernovae have been observed in the
last fifty years (e.g. Barbon et al. 1989, Treffers 1993, Dimai 2005) 
and the Star Formation Rate (SFR) is estimated to be 
SFR=9.5 M$_{\odot}$/yr 
based on it high H$\alpha$ luminosity 
(Kennicutt, 1983).
Values between 5 and 15  M$_{\odot}$/yr are found from different
measurements, mostly H$\alpha$ and FIR 
(e.g. Kennicutt et al. 1994, James et al. 2004,  Sanders et al. 2003).
A number of HII regions is identified in the galaxy,
spread all over the area (Hodge \& Kennicutt 1983, Davis et al. 1997).

Using XMM-Newton data, Davis \& Mushotzky (2004) identify the brightest
X-ray source at the western edge of the galaxy with a ULX with 
luminosity of  L$_{X}^{0.5-10 keV} = 1.1 \times  10^{41}$ erg/s,
making it one of the most luminous known in its class. 
The nuclear source instead is dimmer by at least a factor of several 
thousands than the ROSAT HRI source seen by Davis et al. (1997) eigth
years earlier. 

The finding of a single ULX is in contrast with expectations from star
formation models that assume that ULXs are the heaviest and brightest
tail of the High Mass X-ray binary population. For instance, using the
model of Mapelli et al. (2010), with a metallicity of Z=0.22 Z$_{\odot}$ 
and a mean value of SFR=10 M$_{\odot}$/yr, we predict the formation of 
almost twenty thousand Black Holes in the galaxy and, as a consequence 
of their evolution, about 10 ULXs.
We recall that the number of ULXs expected depends linearly on SFR
and therefore it is uncertain by about a factor of 2 based on the spread
of measured SFR values.

Recent Chandra data (Rasmussen et al. 2006) reveal a shock-like feature 
along the western edge of the
galaxy and a low surface brightness tail extending to the east, similar 
to the morphology seen in other wavebands. Spatially resolved spectroscopy 
shows that the data are consistent with intragroup gas being pressurized at 
the leading western edge of NGC~2276 due to the galaxy moving supersonically 
through the IGM at a velocity of $\sim 850$ km/s.

\begin{figure*}[!t]
\includegraphics[width=17.1cm]{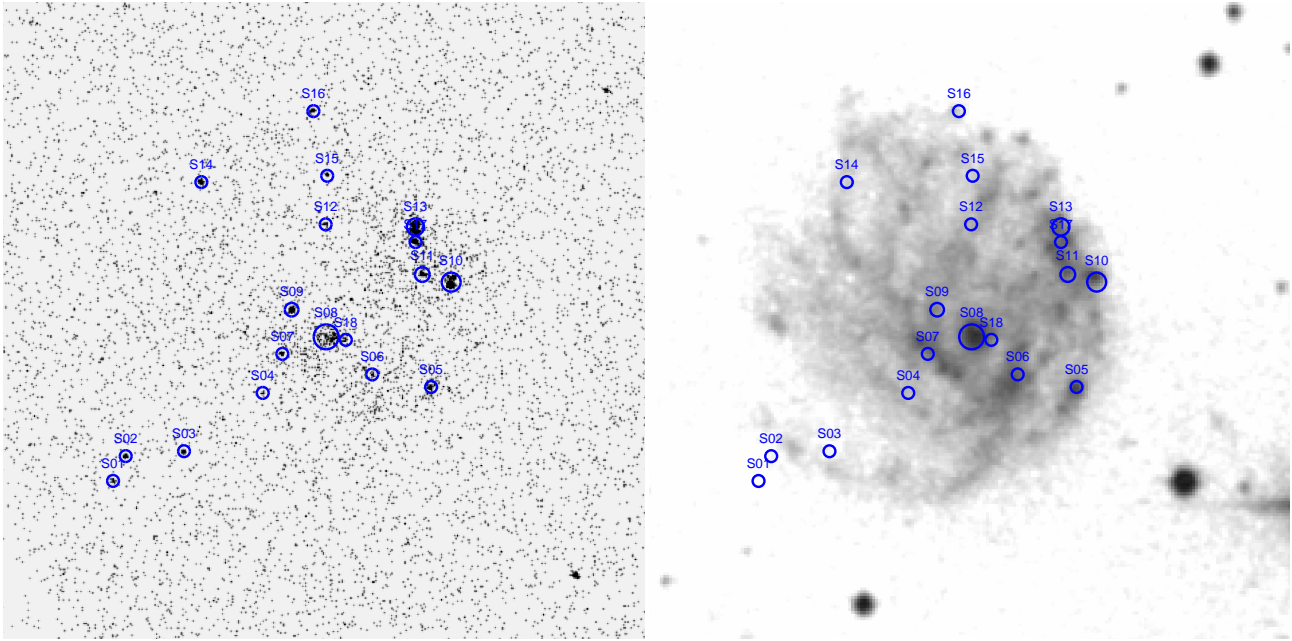}
\caption{{\it Left:} X-ray data from Chandra in the 0.5-7 keV band.
Open circles and numbers identify detected sources.
{\it Right:} The X-ray sources' positions  are overplotted on the 
optical image of NGC~2276 taken from the DSS.}
\label{xray}
\end{figure*}

The diffuse hot gas in NGC~2276 has an X-ray luminosity of 
L$_X^{(0.3-2.0 keV)} = 1.86 \times 10^{40}$ erg/s and a temperature of
$kT \sim 0.35$ keV, with a residual component of unresolved X-ray
binaries of about 10-15\% in flux (Rasmussen et al. 2006).
The IGM measured by Chandra has a profile consistent with previous
measures by Davis et al. (1996) and temperature of kT = 0.85 keV 
and Z= 0.17 Z$_{\odot}$ (Rasmussen et al. 2006).
However, the superb Chandra resolution allows the detection
of a number of point sources not seen or resolved before. The study of the 
population of point sources in the galaxy NGC~2276 and preliminary results 
on the ULXs are the subject of this work.

We assume here a distance of 32.8 Mpc (from NED) an rescale published values
to this distance when necessary.

\section{Chandra point sources}

The Chandra observation of 45.8 ksec has been performed on June 23rd, 2004.
We apply standard reduction procedures and run a sliding cell detection 
algorithm that finds 17 sources in the 
galaxy area. We illustrate the X-ray data and compare with the optical image
of the galaxy in Figure~\ref{xray}.
One of the sources (S08) corresponds to the nuclear region, which had been 
detected previously as variable (Davis \& Mushotzky 2004), 
at L$_X \sim 10^{39}$ erg/s. However, a correct comparison of 
the flux would need a detailed modeling of the sourrounding emission, given
the different resolution of the instrument that have observed NGC~2276
through the years. 
The spectral information is not sufficient to discriminate between a
nuclear active source at very low luminosity or the natural enhancement
at the center of the galaxy. We do not consider this source in the 
construction of the X-ray Luminosity Function (XLF).

Contamination from background sources is estimated according to 
the logN-logS of Hasinger et al. (1993). At the detection limit 
of f$_X^{(0.5-2 keV)} = 7 \times 10^{-16}$ erg cm$^{-2}$ s$^{-1}$ 
(which corresponds
to L$_X = 10^{38}$ erg/s at the distance of NGC~2276) we
expect 1.4 sources by chance in the total area covered by the galaxy. 
To account for this we randomly exclude
one of the sources in the fainter flux range to construct the XLF.

\begin{figure}[htb]
\hbox{
\includegraphics[width=8.3cm]{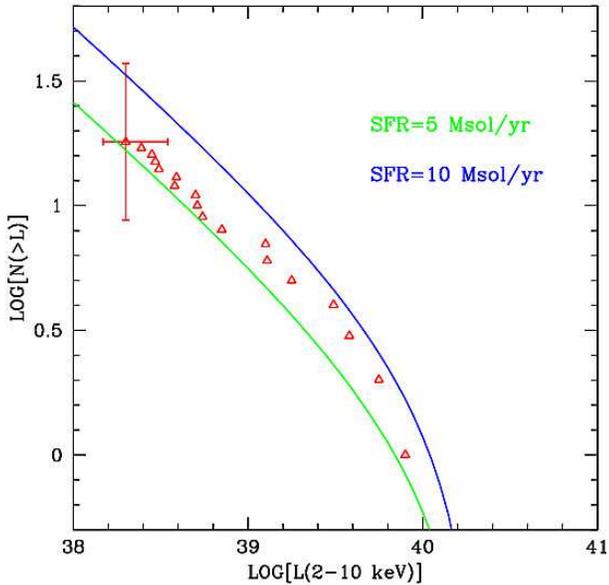} 
}
\caption{The X-ray Luminosity Function computed in the 2-10 keV
band by using the 15 sources
detected in the galaxy area: the nucleus has been removed, as well
as one of the faint sources to account for background contamination.
The solid lines describes the Grimm et al. 2003 luminosity
function for the HMXB, normalized
respectively to 5 and 10 M$_{\odot}$/yr. No formal fit has been performed.
The errorbar on the faintest point are indicative of the uncertainty
in flux and density. }
\label{xlf}
\end{figure}

The XLF is matched to the Universal Luminosity function of High Mass
X-ray binaries (HMXB) proposed by Grimm et al. (2003). The scale parameter 
is the SFR, that we can derive from comparison
with the data. If the model holds, than we see from Figure~\ref{xlf}
that values between 5 and 10 M$_{\odot}$/yr are consistent with
the derived luminosity function. Also comparing the total X-ray luminosity
or the number of ULX with the results of Grimm et al. (2003) we derive
similar results of SFR = 5 and 5.5 M$_{\odot}$/yr respectively.
These values are in the same range of star formation rates derived
in other wavebands (see Section~\ref{Intro}). 
This makes the XLF of NGC~2276 very close to those of the Antennae 
(Zezas et al. 2002) and the Cartwheel (Wolter \& Trinchieri 2004) which are fitted 
respectively with SFR=7.1 M$_{\odot}$/yr and SFR$\sim 20$ M$_{\odot}$/yr.

\section{The ULXs}

\begin{figure}[htb]
\hbox{
\includegraphics[width=8.3cm]{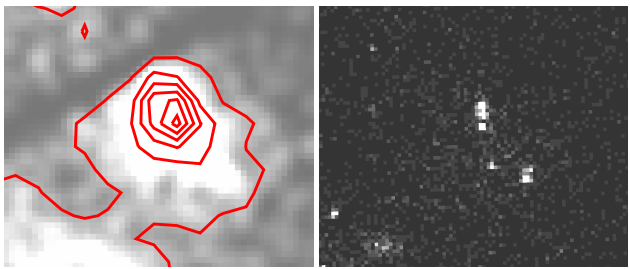} 
}
\caption{{\it Left:} A zoomed in region of the XMM-Newton data representing
the ULX region.  {\it Right:} Chandra data for the same region of the galaxy:
a number of point sources is evident. }
\label{ulx}
\end{figure}

The number of sources with L$_X \geq 10^{39}$ erg/s, which
is the assumed definition for ULXs, are 7, many more than the only ULX 
detected earlier in XMM-Newton data (Davis et al. 1994), and consistent
with the prediction of the model by Mapelli et al. (2010) for a 
SFR = 5-10 M$_{\odot}$/yr.
A comparison of XMM-Newton and Chandra data in the region of the XMM-Newton 
ULX is shown in Fig.\ref{ulx}.  The XMM-Newton source is resolved in 6 
individual sources, of luminosities ranging from a few 
10$^{38}$ erg/s to almost 10$^{40}$ erg/s.
Lower background per detection cell in Chandra also contributes to the 
finding of a large number of point sources in the whole galaxy. 

The brightest sources have enough counts (100-400) to allow a spectral fit:
a single absorbed power law is a good description of these spectra, with 
N$_H \sim 2 \times 10^{21}$ cm$^{-2}$ and 
slopes between $\Gamma = 1.4$ and 2.2, consistent with mean results for
the sample of ULXs studied by Swartz et al. (2004).

To investigate possible variability of the ULX source previously
detected by XMM, we estimate the Chandra luminosity of the
corresponding source as the sum of all the resolved point sources.
We rescale the energy range given by Davis \& Mushotzky (2004) from the
0.5-10 keV to the 2-10 keV band and also correct to the adopted distance
of 32.8 Mpc. This results in a Chandra luminosity of
L$_X^{(2-10keV)} = 2 \times 10^{40}$ erg/s to be
compared to the total luminosity measured by XMM of 
L$_X^{(2-10keV)} = 3.8 \times 10^{40}$ erg/s.
The XMM-Newton value might be affected also by a contribution of diffuse 
gas which falls in the detection cell. However the fraction of hot
gas emission amounts to only about 30\% in the Chandra data. Therefore
it is possible that at least the brightest ULXs have partially dimmed
in the two years between the two observations. We know that variablity
is a common property of ULXs (see e.g. Crivellari et al. 2009 and 
Zezas et al. 2006).
New Chandra observations are needed in order to study the 
variability pattern in detail.


\begin{figure}[!htb]
\hbox{
\includegraphics[width=8.3cm]{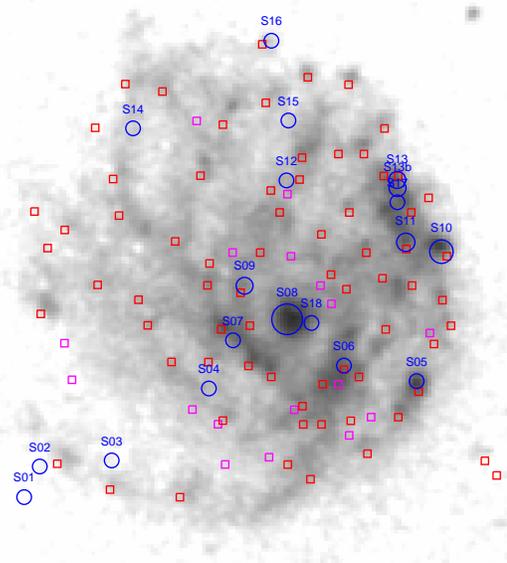} 
}
\caption{The DSS image of the galaxy: overplotted in blue are the
position of the detected Chandra X-ray sources presented in this
work, in red and in magenta are the locations of HII regions listed
in Hodge \& Kennicutt, (1983) Davis et al. (1997) respectively.
The ULXs reported in this paper are S09, S10, S11, S13, S13b, S14, S17.}
\label{hii}
\end{figure}

Hodge et al. (1983) list 72 HII regions detected in the galaxy, and
Davis et al. (1997) find a few more.
The positions of the HII region are plotted in Fig.~\ref{hii}, together
with the position of X-ray sources.
Davis et al. (1997) notice that many of the HII regions are 
also associated with strong non-thermal radio sources, which is a fact 
not well explained, and propose an association with a supernova 
remnant (SNR) in dense environment to explain the emission. 
Although the spatial resolution at the distance of NGC~2276 is not enough
to warrant a physical association, we find that six of the HII region
are positionally coincident ($1^{\prime\prime} - 2^{\prime\prime}$)
also with bright X-ray sources, four of which are ULXs. 
It is tempting therefore to associate the HII regions to the ULXs.
The main interpretation of ULXs is in term of accretion, however,
at least
25\% of all known ULXs have a spatial coincidence with a Supernova (SN), 
and also
one of the brightest known ULXs, N10 in the Cartwheel galaxy, might be
explained as a very bright and peculiar SN, as well as an almost ordinary
HMXB (see Pizzolato et al. 2010).
Therefore a deeper study of the association with HII regions and radio
emission could give more insight in the explanation of ULXs as a class.

\section{Conclusions}

We have analyzed a Chandra observation of NGC~2276, a spiral galaxy in the
group of the elliptical NGC~2300. We have paid particular attention to the 
point sources
detected in the area of the galaxy. We find that a large number of ULXs are
present, which makes NGC~2276 one of the richest enviroment of this
still enigmatic bright sources. Further observations are needed in order
to study in detail spectral and variability properties of this class.

\acknowledgements
We thank the referee for comments.
This work has  made use of the NED database and of the Digital Sky Survey.
We acknowledge partial financial support from INAF through grant PRIN-2007-26.

\vskip 1cm



\begin{thebibliography}{}
  \bibitem{} Barbon, R., Cappellaro, E., Turatto, M., 1989: A\&AS 81, 421
  \bibitem{} Crivellari, E., Wolter, A., \& Trinchieri, G., 2009: A\&A 501, 445
  \bibitem{} Condon, J.J., 1983: ApJS 53, 459
  \bibitem{} Davis, D.S. \& Mushotzky, R.F. 2004: ApJ 604, 653
  \bibitem{} Davis, D.S. Mulchaey, J.S., Mushotzky, R.F., 
Burstein, D., 1996: ApJ 460, 601
  \bibitem{} Davis, D.S., Keel, W.C., Mulchaey, J.S., 
Henning, P.A., 1997: AJ 114, 613
  \bibitem{} Dimai, A. 2005: IAUC 8588, 2
  \bibitem{} Elmegreen, D.M., Sundin, M., Sundelius, B., Elmegreen, B, 1991:
A\&A 244, 52
  \bibitem{} Grimm, H.-J., Gilfanov, M., Sunyaev, R., 2003: MNRAS 339, 793
  \bibitem{} Gruendl, R.A., Vogel, S.N., Davis, D.S., Mulchaey, J.S, 
1993: ApJ 413, L81
  \bibitem{} Hasinger, G., et al. 1993: A\&A 275, 1
  \bibitem{} Hodge P.W. \& Kennicutt, R.C., 1983: AJ 88, 296
  \bibitem{} James, P.A., et al. 2004: A\&A 414, 23
  \bibitem{} Kennicutt, R.C., 1983: ApJ 272, 54
  \bibitem{} Kennicutt, R.C., Tamblyn, P., Congdon C.E. 1994, ApJ 435, 22
  \bibitem{} Mapelli, M., Ripamonti, E., Zampieri, L., Colpi, M.,
Bressan, A., 2010: MNRAS, in press, arXiv:1005.3548
  \bibitem{} Mulchaey, J.S., Davis, D.S., Mushotzky, R.F., \&
Burstein, D.: 1993, ApJ 404, L9
  \bibitem{} Pizzolato, F., Wolter, A., Trinchieri, G., 2010: MNRAS 406, 1116
  \bibitem{} Rasmussen, J., Ponman, T.J., \& Mulchaey, J.S., 2006: MNRAS 370, 453
  \bibitem{} Sanders, D.B., et al., 2003: AJ 126, 1607
  \bibitem{} Swartz, D.A., Ghosh, K.K., Tennant, A.F., Wu, K. 2004:
ApJS 154, 519
  \bibitem{} Treffers, R.R. et al: 1993, IAUC 5850
  \bibitem{} Wolter, A., \& Trinchieri, G., 2004: A\&A 426, 787
  \bibitem{} Zezas, A., \& Fabbiano, G., 2002: ApJ 577, 726
  \bibitem{} Zezas, A., et al. 2006: ApJS 166, 211
\end{thebibliography}
\end{document}